\documentclass[english,aip,jcp]{revtex4-1}
\usepackage[T1]{fontenc}
\usepackage[latin9]{inputenc}
\usepackage[a4paper]{geometry}
\usepackage{babel}
\usepackage{amsmath}
\usepackage{graphicx}
\usepackage[unicode=true]
 {hyperref}

\makeatletter

\providecommand{\tabularnewline}{\\}

\providecommand{\tabularnewline}{\\}

\makeatother

\begin{document}

\title{The parameters uncertainty inflation fallacy}

\author{Pascal PERNOT}

\affiliation{Laboratoire de Chimie Physique, UMR8000, CNRS / Univ. Paris-Sud,
F-91405 Orsay, France}
\email{pascal.pernot@u-psud.fr}

\begin{abstract}
\noindent Statistical estimation of the prediction uncertainty of
physical models is typically hindered by the inadequacy of these models
due to various approximations they are built upon. The prediction
errors caused by model inadequacy can be handled either by correcting
the model's results, or by adapting the model's parameters uncertainty
to generate prediction uncertainties representative, in a way to be
defined, of model inadequacy errors. The main advantage of the latter
approach (thereafter called PUI, for Parameters Uncertainty Inflation)
is its transferability to the prediction of other quantities of interest
based on the same parameters. A critical review of implementations
of PUI in several areas of computational chemistry shows that it is
biased, in the sense that it does not produce prediction uncertainty
bands conforming with model inadequacy errors. 
\end{abstract}
\maketitle

\section{Introduction}

Prediction uncertainty of physical models or simulations is difficult
to estimate \cite{Glotzer2009}. Yet, it is a necessary step to produce
virtual measurements \cite{Irikura2004}, \emph{i.e.}, to enable simulations
or models to replace experiments.

Estimation of model prediction uncertainty requires a thorough analysis
of three major error sources: (i) systematic errors due to the model
formulation and approximations (model inadequacy); (ii) numerical
errors (notably for stochastic models); and (iii) parameter uncertainty.
Numerical errors are expected to be kept to a negligible or well controlled
level (except maybe for chaotic model) \cite{Irikura2004,Williams2008,Feher2012},
while parameter uncertainty is estimated by well established calibration
methods, notably bayesian inference \cite{Gregory2005,Gelman2013,McElreath2015}.
The most challenging part of the uncertainty quantification process
remains model inadequacy \cite{OHagan2013}, which takes often a major
fraction of the uncertainty budget \cite{Irikura2007}.

\begin{figure}[!t]
\noindent \begin{centering}
\includegraphics[width=0.45\textwidth]{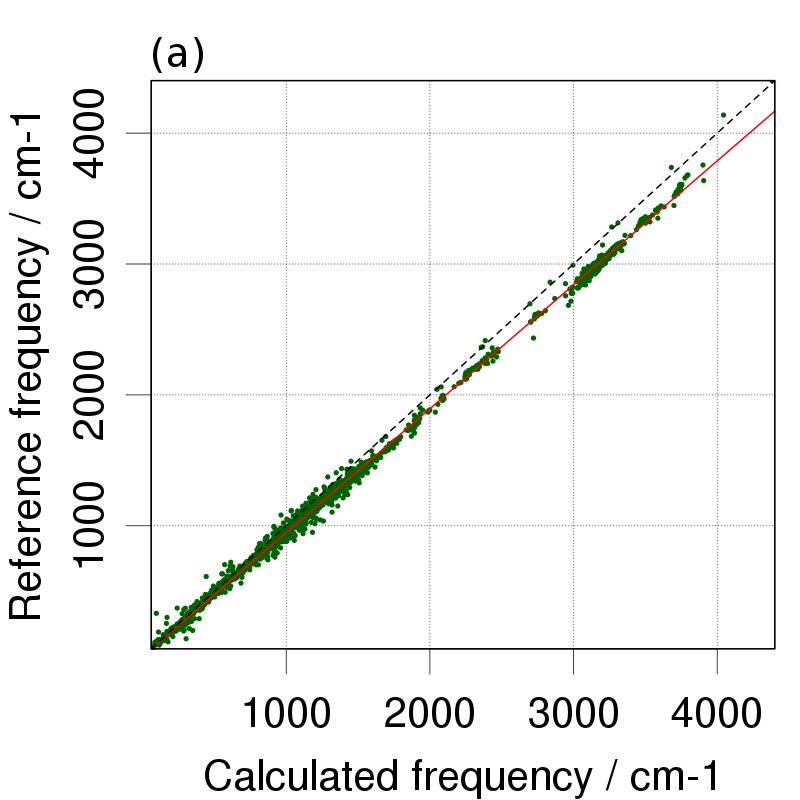} \includegraphics[width=0.45\textwidth]{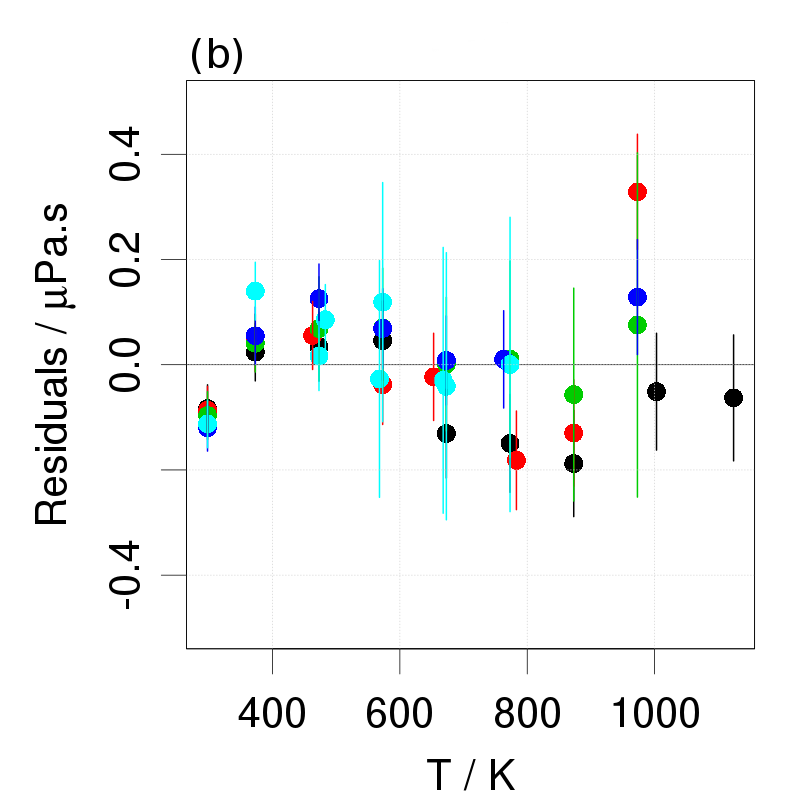} 
\par\end{centering}
\caption{\label{fig:Inadequacy}Examples of model inadequacy: \protect \protect \\
 (a) scatterplot of reference fundamental vibrational frequencies
with respect to harmonic \emph{ab initio} frequencies calculated at
the CCD/6-31G{*} theory/basis-set level. The red line depicts the
linear tendency in the data cloud, which is not the unit line. The
error bars on the reference data are invisible at this scale. The
data are extracted from the CCCBDB \cite{cccbdb_6_1};\protect \protect \\
 (b) residuals of the fit of Argon viscosity data by a Chapman-Enskog
model. The error bars represent 2-$\sigma$ experimental confidence
intervals. Even if the empirical model achieves well centered residuals,
the T-dependent oscillation of the latter reveals an unsatisfactory
fit, notably at low temperature. This dataset is described by Cailliez
and Pernot \cite{Cailliez2011}.}
\end{figure}

Model inadequacy is characterized by the inability of a model to produce
results in statistical agreement with reference data, within their
uncertainty range. Even empirical physical models, having adjustable
parameters, cannot always achieve a statistically valid representation
of the reference data used for their calibration. As model improvement
is often impractical or impossible, it is important to be able to
deal with the limitations of existing models. Model inadequacy should
not be seen as a failure of physical models, but more as an intrinsic
component of their predictions that has to be taken care of. Two examples
are provided and commented in Fig~\ref{fig:Inadequacy}.

Prediction errors due to model inadequacy can be handled either \emph{internally},
by model improvement in the spirit of Jacob's ladder for DFT \cite{Perdew2001}
and composite methods of quantum chemistry \cite{Raghavachari2015},
or \emph{externally}, by statistical correction of model predictions.
The focus of the present study is on the latter approach, which consists
in designing a statistical model representing the unexplained part
of the model residuals on a set of reference data \cite{Kennedy2001,Santner2003,Lejaeghere2014a,Pernot2015,DeWaele2016,Proppe2016,Proppe2017}.

A major drawback of the statistical correction of model predictions
is its lack of transferability to other observables \cite{Campbell2006,Oliver2015},
which is an issue with generalist models, such as atomistic/molecular
simulation or electronic structure computing. As model parameters
and their uncertainties are in principle transferable, a solution
is to assign them the residual dispersion due to model inadequacy,
by a controlled increase in parameters uncertainty. This has been
implemented, for instance, through ensemble methods in the calibration
of density functionals approximations \cite{Mortensen2005,Wellendorff2012,Lejaeghere2013,Wellendorff2014,Pernot2015,Pandey2015,Lejaeghere2016,Simm2016},
or through the concept of embedded models \cite{Sargsyan2015}.

However, this \emph{parameter uncertainty inflation} (PUI) approach
suffers from intrinsic limitations which have to be carefully considered: 
\begin{enumerate}
\item Due to the geometry of the problem in data space \cite{Transtrum2011,Pernot2017},
enlarging the uncertainty patch on the model manifold around the optimal
parameters does not contribute to improve the validity of an inadequate
model. Besides, even if model adequacy were recovered by PUI for a
calibration property, no guarantee exists on the transferability of
adequacy to other properties of interest, which have different model
manifolds.
\item Considering a model $M(x;\boldsymbol{\vartheta})$,\footnote{Boldface type refers to vectors or matrices.}
depending on a control variable $x$ (\emph{e.g.} temperature, pressure...),
and parameters $\boldsymbol{\vartheta}$, propagation of parameter
uncertainty is governed by the functional shape of the model sensitivity
coefficients $(\partial M(x;\boldsymbol{\vartheta})/\partial\vartheta_{i})$
as functions of $x$ \cite{GUM}. This means that the shape of the
prediction uncertainty bands over the control space does not necessarily
conform with the shape of the model inadequacy errors. As will be
shown below, this might lead to uncontrolled under- or over-estimation
of prediction uncertainty, depending on the value of the control variables. 
\end{enumerate}
This short study focuses on the second problem and considers a series
of examples inspired from the computational chemistry literature.
It focuses on deterministic models, or stochastic models with negligible
numerical errors. The next section introduces three methods implementing
the PUI approach in a common bayesian framework. Section \ref{sec:Examples}
treats three examples: (1) a simple linear model involving the statistical
correction of \emph{ab initio} molecular vibrational frequencies;
(2) a meta-analysis of the prediction uncertainty for formation heats
of solids calculated by the mBEEF density functional; and (3) an original
application to the calibration of a Lennard-Jones potential on temperature-dependent
viscosity data. A discussion of the encountered problems and recommendations
to users of these PUI methods serve as conclusion in Section \ref{sec:Discussion-and-Conclusion}.

\section{Methods\label{sec:Methods}}

Bayesian data analysis is a convenient framework to develop calibration-prediction
methods, and it has been used here to present and develop PUI methods.
A brief introduction to bayesian analysis is provided in the next
section. More details can be found in several excellent textbooks
\cite{Gregory2005,Gelman2013,McElreath2015}.

\subsection{Statistical calibration and prediction}

One considers a model represented by the function $M(x;\boldsymbol{\vartheta})$,
which parameters $\boldsymbol{\vartheta}$ have to be identified,
\emph{i.e.} characterized by their probability density function (pdf)
or, in the gaussian hypothesis, their ``best'' value and covariance
matrix.

\paragraph{Calibration.}

Parameters inference is done by calibration of the model on a set
of reference data $\boldsymbol{D}=\left\{ x_{i},y_{i}\right\} _{i=1}^{N}$,
accompanied by uncertainties $\left\{ u_{y_{i}}\right\} _{i=1}^{N}$.
In the general case, the full covariance matrix, $\boldsymbol{V}_{D}$,
might be available in addition to the usual diagonal elements ($u_{y_{i}}$).
All the knowledge about the parameters is encoded in the \emph{posterior}
pdf $p(\boldsymbol{\vartheta}|\boldsymbol{D})$ for $\boldsymbol{\vartheta}$,
conditional on $\boldsymbol{D}$ (and $M$). The posterior pdf is
obtained by Bayes theorem 
\begin{equation}
p(\boldsymbol{\vartheta}|\boldsymbol{D})\propto p(\boldsymbol{D}|\boldsymbol{\vartheta})\,p(\boldsymbol{\vartheta}),
\end{equation}
where $p(\boldsymbol{\vartheta})$ is the prior pdf of the parameters
and $p(\boldsymbol{D}|\boldsymbol{\vartheta})$ is the likelihood.
Assuming normal data error distribution, the likelihood can be written
as 
\begin{equation}
p(\boldsymbol{D}|\boldsymbol{\vartheta})\propto\left|\boldsymbol{V}_{D}\right|^{-1/2}\exp\left(-\frac{1}{2}\boldsymbol{R}^{T}\boldsymbol{V}_{D}^{-1}\boldsymbol{R}\right),\label{eq:likelihood}
\end{equation}
where $\boldsymbol{R}$ is the column vector of residuals 
\begin{equation}
R_{i}(\boldsymbol{\vartheta})=y_{i}-M(x_{i};\boldsymbol{\vartheta}).
\end{equation}

The maximum a posteriori (MAP) 
\begin{equation}
\boldsymbol{\hat{\vartheta}}=\mathrm{argmax}_{\boldsymbol{\vartheta}}p(\boldsymbol{\vartheta}|\boldsymbol{D})
\end{equation}
is a point estimate of the set of parameters providing the best fit
to the data, constrained by the prior pdf. The mean value of the parameters
$\boldsymbol{\mu}_{\vartheta|D}$ and their covariance matrix $\boldsymbol{V}_{\vartheta|D}$
are often used to summarize the posterior pdf. It is important to
note that, except if the model is not identifiable, the variance of
the parameters is a decreasing function of the calibration dataset
cardinal. Moreover, the covariance matrix of the parameters should
not be used if the model calibration is not statistically valid.

Validation of a calibration can be done by posterior predictive assessment
(see below) \cite{Vehtari2012,Gelman2013,Oliver2015}, but simple
statistics, such as the Birge ratio \cite{Birge1932,Kacker2008} can
be very useful. It is defined as 
\begin{equation}
R_{B}=\frac{1}{N-n}\boldsymbol{R}^{T}(\boldsymbol{\hat{\vartheta}})\boldsymbol{V}_{D}^{-1}\boldsymbol{R}(\boldsymbol{\hat{\vartheta}}),\label{eq:Birge}
\end{equation}
where $n$ is the number of parameters in the model, and should be
close to 1 for satisfactory fits. Values smaller than 1 point to over-estimated
data variance, while too high values can be due to under-estimated
data variance or, most often, to model inadequacy.

\paragraph{Prediction.}

For deterministic models, the mean value of a prediction at a new
control value $\tilde{x}$ and its variance can be approximated by
linear uncertainty propagation \cite{GUM} 
\begin{align}
\mu_{M|D}(\tilde{x}) & =M(\tilde{x};\boldsymbol{\mu}_{\vartheta|D})\label{eq:LUP-mean}\\
u_{M|D}^{2}(\tilde{x}) & =\boldsymbol{J}^{T}(\tilde{x};\boldsymbol{\mu}_{\vartheta|D})\boldsymbol{V}_{\vartheta|D}\boldsymbol{J}(\tilde{x};\boldsymbol{\mu}_{\vartheta|D}),\label{eq:LUP-var}
\end{align}
where $\boldsymbol{J}$ is a vector of \emph{sensitivity coefficients}
\begin{equation}
\boldsymbol{J}_{k}(x;\boldsymbol{\mu}_{\vartheta|D})=\left.\frac{\partial M(x;\boldsymbol{\vartheta})}{\partial\boldsymbol{\vartheta}_{k}}\right|_{\boldsymbol{\mu}_{\vartheta|D}}.\label{eq:Sensitivity}
\end{equation}
The linear approximation is exposed here mostly for didactic reasons.
If it is not appropriate, one has to estimate $\mu_{M|D}$ and $u_{M|D}^{2}$
by higher order Taylor expansions \cite{GUM}, or by numerical integration
(Monte Carlo method) \cite{GUMSupp1}.

Various prediction statistics can be used for model validation \cite{Vehtari2012,Gelman2013,Oliver2015}.
Posterior predictive assessment compares model predictions with reference
data and/or validation data. Visual inspection of prediction probability
intervals (prediction bands) is generally very useful.

In the following, one will mostly refer to the mean prediction variance
on the calibration set 
\begin{align}
MPV & =\frac{1}{N}\sum_{i=1}^{N}u_{M|D}^{2}(x_{i}),
\end{align}
and note the mean prediction uncertainty as 
\begin{equation}
\overline{u}_{M|D}=\sqrt{MPV}.
\end{equation}
The mean squared errors of the model at the MAP 
\begin{equation}
MSE=\frac{1}{N}\sum_{i=1}^{N}R_{i}^{2}(\boldsymbol{\hat{\vartheta}})
\end{equation}
will be used as a reference point for the validation of model predictions.

\paragraph{Inadequate models.}

If the covariance matrix of the reference dataset used for model validation
is known, a Birge ratio value higher than 1 is a good indicator of
model inadequacy. Otherwise, inspection of the residuals and comparison
with typical reference data uncertainties is often used. A notable
trend in the residuals, possibly quantified by their correlation length,
is also a feature to be checked.

\subsection{The Parameters Uncertainty Inflation strategy}

The aim of PUI is to adjust a model's parameters uncertainty in order
to produce enough model output variance to encompass the part of the
variance in the residuals due to model inadequacy. This is achieved
in Eq.~\ref{eq:LUP-var} by adapting the parameters covariance matrix
$\boldsymbol{V}_{\vartheta|D}$. Two methods are considered: an \emph{indirect}
one, based on a scaling of the data covariance matrix $\boldsymbol{V}_{D}$
(Eq.~\ref{eq:likelihood}); and a \emph{direct} one, based on the
optimization of the elements of $\boldsymbol{V}_{\vartheta|D}$, a
more complex option with several variants.

\subsubsection{The indirect approach\label{par:Variance-inflation}}

A statistical approach, inspired from bayesian statistics, developed
by Brown and Sethna \cite{Brown2003}, and adapted by Frederiksen
\emph{et al.} \cite{Frederiksen2004,Mortensen2005} identifies ``parameters
ensembles'' from which prediction statistics are estimated. To relieve
the problem of model inadequacy, a scaling factor, $T$, is introduced
in the pdf describing the ensemble.

Translating this in the bayesian framework, an \emph{empirical} likelihood
is used 
\begin{equation}
p(\boldsymbol{D}|\boldsymbol{\vartheta},T)\propto\left|T\boldsymbol{V}_{D}\right|^{-1/2}\exp\left(-\frac{1}{2T}\boldsymbol{R}^{T}\boldsymbol{V}_{D}^{-1}\boldsymbol{R}\right),\label{eq:likelihood-1}
\end{equation}
which can be seen as the standard likelihood (Eq.~\ref{eq:likelihood})
with a scaled data covariance matrix $T\boldsymbol{V}_{D}$.

Jacobsen and collaborators choose $T$ so that the mean variance of
model predictions reproduces the mean squared error for the best parameters
\cite{Wellendorff2014}, \emph{i.e.} 
\begin{equation}
MPV(T)\simeq MSE.\label{eq:indirect}
\end{equation}
This equation assumes that model inadequacy is a strongly dominant
part of the residuals, otherwise, data uncertainty should be explicitly
considered. In the ensemble method, $T$ is chosen using a statistical
mechanics analogy with a temperature, leading to \cite{Frederiksen2004}
\begin{equation}
T=\frac{2C_{0}}{n},
\end{equation}
where $C_{0}=\frac{1}{2}\boldsymbol{R}^{T}(\boldsymbol{\hat{\vartheta}})\boldsymbol{V}_{D}^{-1}\boldsymbol{R}(\boldsymbol{\hat{\vartheta}})$,
and $n$ is the number of parameters.

It is thorough to establish the link with the Birge ratio, using Eq.~\ref{eq:Birge},
as 
\begin{equation}
T=\frac{N-n}{n}\,R_{B}.\label{eq:T-RB}
\end{equation}
Note that this \emph{indirect} PUI method is akin to the Birge ratio
method used in metrological inter-laboratory comparisons to reconcile
inconsistent data \cite{Kacker2008,Bodnar2014}. The Birge ratio method,
assuming an adequate model and misestimated data variances, rescales
the latter in order to get a valid statistical estimation of the data
mean, whereas, in the hypothesis of reliable data variances, $T$
is chosen here to compensate for model inadequacy and obtain valid
prediction statistics.

An alternative estimation of $T$ can be based on Eqns.~\ref{eq:LUP-var}
and \ref{eq:indirect}, assuming a near-linear dependence of the model
on its parameters in their uncertainty range and negligible data uncertainty:
\begin{equation}
T\simeq\frac{MSE}{MPV(T_{0})},\label{eq:varinf}
\end{equation}
using the mean prediction variance from a reference calibration with
$T=T_{0}\equiv1$\emph{.}

\subsubsection{The direct approach\label{subsec:Direct-approach}}

In the direct approach, the model's parameters are considered as random
variables, with a pdf conditioned by a set of hyperparameters, typically
their mean values $\boldsymbol{\mu}_{\vartheta}$ and a covariance
matrix $\boldsymbol{V}_{\vartheta}$, defining a multivariate normal
distribution $p(\boldsymbol{\vartheta}|\boldsymbol{\mu}_{\vartheta},\boldsymbol{V}_{\vartheta})$.

Such stochastic parameters can be handled in the bayesian inference
problem, either at the model level, leading to use a stochastic model
within the standard likelihood framework (Eq.~\ref{eq:likelihood}),
or at the likelihood level.

\paragraph{Model level.\label{par:Model-level}}

At the model level, one estimates the impact of stochastic parameters
on model predictions by uncertainty propagation \cite{GUMSupp1} 
\begin{equation}
f_{M}(\boldsymbol{\xi};\boldsymbol{x},\boldsymbol{\mu}_{\vartheta},\boldsymbol{V}_{\vartheta})=\int\thinspace\prod_{i=1}^{N}\delta\left(\xi_{i}-M(x_{i};\boldsymbol{\vartheta})\right)\,p(\boldsymbol{\vartheta}|\boldsymbol{\mu}_{\vartheta},\boldsymbol{V}_{\vartheta})\,d\boldsymbol{\vartheta},
\end{equation}
where $f_{M}(.;\boldsymbol{x},\boldsymbol{\mu}_{\vartheta},\boldsymbol{V}_{\vartheta})$
is the multivariate pdf of the model's predictions at the vector of
control points $\boldsymbol{x}$. Inserting this stochastic model
in Eq.~\ref{eq:likelihood} can be done by replacing $M(x_{i};\boldsymbol{\vartheta})$
by the mean predictions (Eq.~\ref{eq:LUP-mean}) and their covariance
matrix $\boldsymbol{V}_{M}$ 
\begin{equation}
p(\boldsymbol{D}|\boldsymbol{\mu}_{\vartheta},\boldsymbol{V}_{\vartheta})\propto\left|\boldsymbol{V}_{D}+\boldsymbol{V}_{M}\right|^{-1/2}\exp\left(-\frac{1}{2}\boldsymbol{R}^{T}\left(\boldsymbol{V}_{D}+\boldsymbol{V}_{M}\right)^{-1}\boldsymbol{R}\right),\label{eq:likelihood-2}
\end{equation}
where 
\begin{align}
\boldsymbol{V}_{M,ij} & \equiv u_{M}^{2}(x_{i},x_{j})=\boldsymbol{J}^{T}(x_{i};\boldsymbol{\mu}_{\vartheta})\boldsymbol{V}_{\vartheta}\boldsymbol{J}(x_{j};\boldsymbol{\mu}_{\vartheta}),\label{eq:VM}\\
R_{i} & =y_{i}-\mu_{M|D}(x_{i}).\label{eq:residMean}
\end{align}

Note that using the full variance matrix of Eq.~\ref{eq:likelihood-2}
in the calculation of the Birge ratio (Eq.~\ref{eq:Birge}), by increasing
the variance without affecting the residuals, should enable to validate
the model with $R_{B}\simeq1$.

For a deterministic model $M$, when the number of parameters is smaller
than the number of data points, $\boldsymbol{V}_{M}$ is singular
(non positive-definite), causing the likelihood to be degenerate,
and the calibration to be intractable \cite{Sargsyan2015}. By definition,
for inadequate models, the data covariance matrix is too small to
alleviate the degeneracy problem.

As all data points cannot be reproduced \emph{simultaneously} by the
model, one has to replace the multivariate problem by a set of univariate
problems (marginal likelihoods \cite{Sargsyan2015}), \emph{i.e.},
one ignores the covariance structure of model predictions by taking
\begin{equation}
\boldsymbol{V}_{M,ij}=u_{M}^{2}(x_{i},x_{j})\delta(i-j).
\end{equation}

\paragraph{Likelihood level.\label{par:Likelihood-level}}

A new likelihood, conditioned on the hyperparameters to be inferred
\cite{Oliver2015,Sargsyan2015}, is obtained by integration of the
standard likelihood (Eq.~\ref{eq:likelihood}) over the possible
values of the parameters (marginalization) 
\begin{equation}
p(\boldsymbol{D}|\boldsymbol{\mu}_{\vartheta},\boldsymbol{V}_{\vartheta})=\int p(\boldsymbol{D}|\boldsymbol{\vartheta})p(\boldsymbol{\vartheta}|\boldsymbol{\mu}_{\vartheta},\boldsymbol{V}_{\vartheta})\,d\boldsymbol{\vartheta}.\label{eq:integLik}
\end{equation}

As in the previous case, it is pointed out by Sargsyan \emph{et al.}
\cite{Sargsyan2015} that this likelihood is in general degenerate,
so that the inference problem has to be solved by alternative methods,
such as Approximate Bayesian Computation (ABC) \cite{Csillery2010,Sunnaker2013}.
In this case, the full likelihood (Eq.~\ref{eq:integLik}) is replaced
by a tractable expression, involving summary statistics of the model
predictions, to be compared to similar statistics of the data. An
example is provided in Sargsyan \emph{et al.} \cite{Sargsyan2015},
where the mean value of the model and its prediction uncertainty are
used. A version adapted to the present problem, with an explicit treatment
of experimental uncertainty is 
\begin{equation}
p_{ABC}(\boldsymbol{D}|\boldsymbol{\mu}_{\vartheta},\boldsymbol{V}_{\vartheta})\propto\exp\left(-\frac{1}{2}\boldsymbol{R}^{T}\boldsymbol{V}_{D}^{-1}\boldsymbol{R}\right)\times p_{reg}(\boldsymbol{D}|\boldsymbol{\mu}_{\vartheta},\boldsymbol{V}_{\vartheta})
\end{equation}
where the first term has the same expression as the standard likelihood
(Eq.~\ref{eq:likelihood}) using residuals evaluated at the mean
of the model prediction (Eq.~\ref{eq:residMean}), and the second
term ensures that the predicted model uncertainty $u_{M}(x_{i})$,
combined with experimental uncertainty $u_{y_{i}}$, is of a magnitude
compatible with the residuals 
\begin{equation}
p_{reg}(\boldsymbol{D}|\boldsymbol{\mu}_{\vartheta},\boldsymbol{V}_{\vartheta})=\exp\left(-\sum_{i=1}^{N}\frac{\left(\sqrt{u_{M}^{2}(x_{i})+u_{y_{i}}^{2}}-|R_{i}|\right){}^{2}}{2u_{y_{i}}^{2}}\right).\label{eq:ABC}
\end{equation}
As evidenced in our notation, this term can also be seen as a regularization
function, necessary to constrain the parameters covariance matrix
$\boldsymbol{V}_{\vartheta}$ in the inference process. The constraint
imposed here is a statistical variant of Eq.~\ref{eq:indirect},
but aims at the same effect.

\section{Applications\label{sec:Examples}}

\subsection{Harmonic vibrational scaling factors\label{sec:Vibrational-scaling-factors}}

Various approximations in the \emph{ab initio} calculation of harmonic
vibrational frequencies of molecules lead to a systematic bias with
respect to fundamental experimental frequencies (Fig.~\ref{fig:Inadequacy}(a)),
which can be statistically corrected by a simple scaling of the calculated
values \cite{Pople1981,DeFrees1985,Rauhut1995,ScoRad-96,Halls2001,Neugebauer2003}.
This \emph{a posteriori} scaling corrects empirically for the approximations
involved in the \emph{ab initio} calculation. After scaling, the residual
errors are typically still much larger than the reference data uncertainties
\cite{Irikura2005}, and the corrected model is still inadequate ($R_{B}\gg1$).
\begin{figure}[!t]
\noindent \begin{centering}
\includegraphics[width=0.9\textwidth]{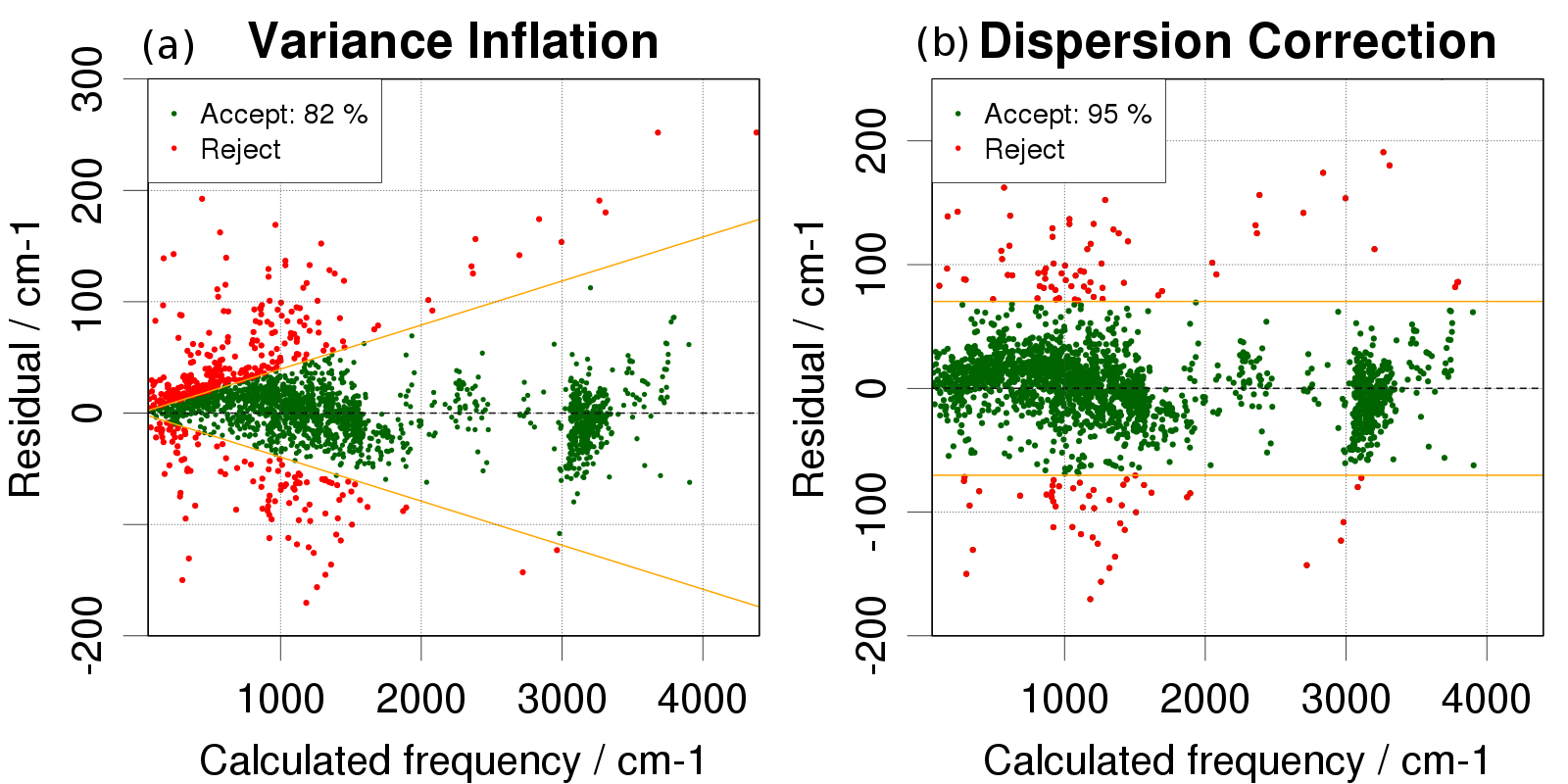} 
\par\end{centering}
\caption{\label{mar:freq_VIF}Scatter plots of the residuals of 2279 scaled
vibrational frequencies for the CCD/6-31G{*} theory/basis-set level.
The orange lines show the 95\,\% confidence prediction range for
two methods: (a) parameters uncertainty inflation; (b) dispersion
correction. The red points lie outside of the 95\,\% prediction range. }
\end{figure}

One considers here the scaling model $M(x;s)=s*x$, where $s$ is
the scaling factor and $x$ a calculated frequency. Irikura \emph{et
al.} \cite{Irikura2005} proposed a method to evaluate the prediction
uncertainty of vibrational frequencies corrected by scaling factors.
Their approach assumes a multiplicative uncertainty model 
\begin{eqnarray}
\nu_{i} & = & s*x_{i},\label{eq:mnui}\\
u_{\nu_{i}} & = & u_{s}*x_{i},\label{eq:unui}
\end{eqnarray}
where $\nu_{i}$ a scaled frequency and $u_{s}$ is the scaling factor
uncertainty. An expression of $u_{s}$ has been derived (Eq. 21 in
Irikura \emph{et al.} \cite{Irikura2005}) as 
\begin{equation}
u_{s}^{2}\simeq\frac{\sum_{i}(y_{i}-s*x_{i})^{2}}{\sum_{i}x_{i}^{2}},\label{eq:us}
\end{equation}
which is different from the uncertainty that would result from an
ordinary least squares calibration model \cite{Pernot2011}, \emph{i.e.},
for large data sets, 
\begin{equation}
u_{s}^{2}\simeq\frac{MSE}{N\sum_{i}x_{i}^{2}}.
\end{equation}

We want to emphasize here that it is possible to recover Eq.~\ref{eq:us}
by the indirect PUI approach. Namely, equating the mean prediction
variance with the mean squared errors (Eq.~\ref{eq:indirect}) leads
to 
\begin{align}
\frac{1}{N}\sum_{i}u_{\nu_{i}}^{2} & =\frac{1}{N}\sum_{i}(y_{i}-s*x_{i})^{2},
\end{align}
while Eq.~\ref{eq:unui} gives 
\begin{align}
\sum_{i}u_{\nu_{i}}^{2} & =u_{s}^{2}\sum_{i}x_{i}^{2},
\end{align}
from which one derives Eq.~\ref{eq:us}.

This shows clearly that the derivation of $u_{s}$ by Irikura \emph{et
al.} \cite{Irikura2005} does not provide the uncertainty on the scale
parameter resulting from the calibration procedure, but the parameter
uncertainty necessary to recover a prediction variance over the calibration
set compatible with the model errors, in the hypothesis of a multiplicative
uncertainty model.

To illustrate the implication of this choice on prediction uncertainty
bands, let us consider a set of vibrational frequencies extracted
from the CCCBDB \cite{cccbdb}. A link to the \texttt{R} \cite{RTeam2015}
scripts used for data extraction, cleanup and treatment is provided
in the Supporting Information section.

\begin{figure}[!t]
\noindent \begin{centering}
\includegraphics[width=0.5\textwidth]{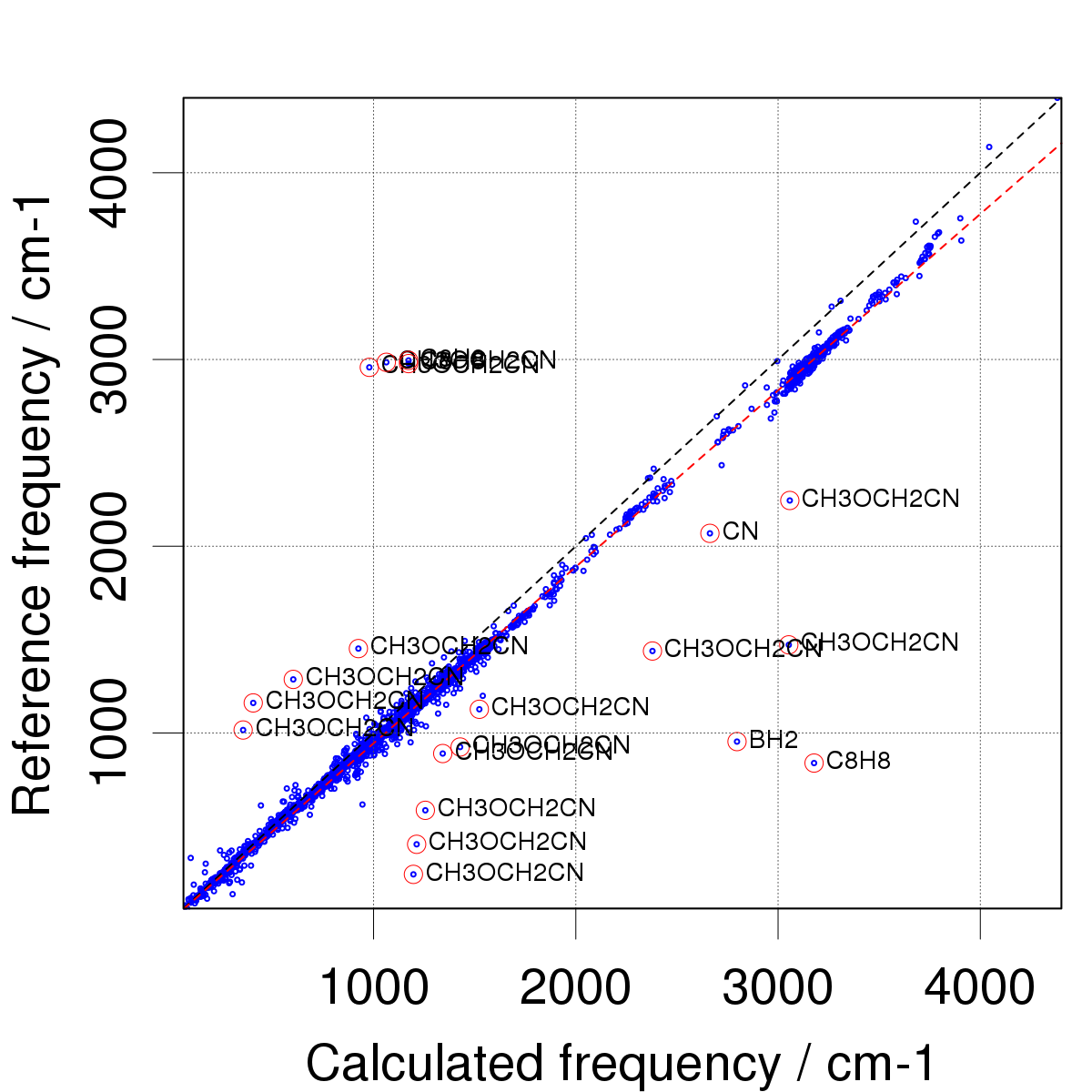} 
\par\end{centering}
\caption{\label{mar:check}Scatterplot of reference fundamental vibrational
frequencies with respect to harmonic \emph{ab initio} frequencies
calculated at the CCD/6-31G{*} theory/basis-set level, as directly
extracted from the CCCBDB \cite{cccbdb_6_1}; the most outlying points
have been circled and labeled.}
\end{figure}

For the CCD/6-31G{*} theory/basis-set combination, a data set containing
2323 frequencies is recovered (7 records with incomplete data have
been removed). A sanity check, based on a plot of the reference frequencies
\emph{vs}. the calculated frequencies, enables to detect several aberrant
points, mainly due to incorrect symmetry assignment for CH$_{3}$OCH$_{2}$CN
and C$_{8}$H$_{8}$ (Fig.~\ref{mar:check}). Also, the CN frequency
and one BH$_{2}$ frequency are outstanding and marked as outliers.
These data were removed, and the final data set contains 2279 frequencies.\footnote{For the improvement of this invaluable database, I strongly encourage
to report any observed data problem to the CCCBDB curator, through
the error form at http://cccbdb.nist.gov/errorformx.asp.} Let us note that a more rigorous data curation procedure would be
required to generate reference scaling factors, which is not the aim
of the present paper.

The statistical analysis of this set gives $s=0.947$, in conformity
with the CCCBDB value \cite{cccbdb_6_1}, and $u_{s}=0.020$ (Eq.~\ref{eq:us}),
smaller than the value of 0.046 reported in the CCCBDB, which reflects
the impact of aberrant points in the original dataset on $u_{s}$.

One can check on Fig.~\ref{mar:freq_VIF}(a) that the linear dependence
of the prediction uncertainty implied by Eq.~\ref{eq:unui} is not
representative of the residuals cloud. It underestimates the dispersion
at low frequency and overestimates it at high frequency. Furthermore,
the probability for a calibration data point to lie in a 95 percent
confidence band $\left[-2u_{s}x_{i},\,2u_{s}x_{i}\right]$ is only
82\,\%.

It has been shown that in this case, model inadequacy should not be
accounted for by Eq.~\ref{eq:unui} \cite{Pernot2010a,Pernot2011}.
Instead, the completion of the model by a stochastic variable $\delta\sim N(0,\sigma^{2})$
representing model inadequacy, 
\begin{eqnarray}
\nu_{i} & = & s*x_{i}+\delta\\
u_{\nu_{i}}^{2} & = & u_{s}^{2}*x_{i}^{2}+\sigma^{2},\label{eq:unui-1}
\end{eqnarray}
enables a more consistent estimation of prediction uncertainty bands
(Fig.~\ref{mar:freq_VIF}(b)). The use of the stochastic correction
$\delta$ recognizes that the model errors have a random and uniform
distribution with respect to the control variable. In this case, $u_{s}=4.1\,10^{-4}$
is the standard uncertainty of the scale factor resulting from ordinary
least-squares regression \cite{Pernot2011}. For large calibration
datasets like the present one, the first term of the prediction variance
is negligible, and one finds that $u_{\nu_{i}}^{2}\simeq\sigma^{2}\simeq MSE$
\cite{Pernot2011,Pernot2015}.

This example shows how the one-parameter scaling model, and the implied
sensitivity coefficient, prevents the indirect PUI strategy to achieve
reasonable prediction uncertainty bands. It is now acknowledged that
the uncertainty factor defined by Eq.~\ref{eq:us} should not be
used for prediction uncertainty \cite{Irikura2011,Jacobsen2013},
although this is not clearly stated in the CCCBDB where the corresponding
values of $u_{s}$ are still reported \cite{cccbdb_6_1}.

\subsection{Calibration of density functional approximations\label{sec:Calibration-of-DFAs}}

Jacobsen and coworkers \cite{Frederiksen2004,Mortensen2005,Petzold2012,Wellendorff2012,Medford2014,Pandey2015}
have elaborated an ensemble method to account for the uncertainty
in the parameters of their calibrated mBEEF density functional. Considering
that the prediction errors are typically much larger than the reference
data uncertainty (model inadequacy), they scale the parameter covariance
matrix to get a mean prediction uncertainty in agreement with the
MSE (Eq.~\ref{eq:indirect}).
\begin{figure}[!t]
\begin{centering}
\includegraphics[width=0.9\textwidth]{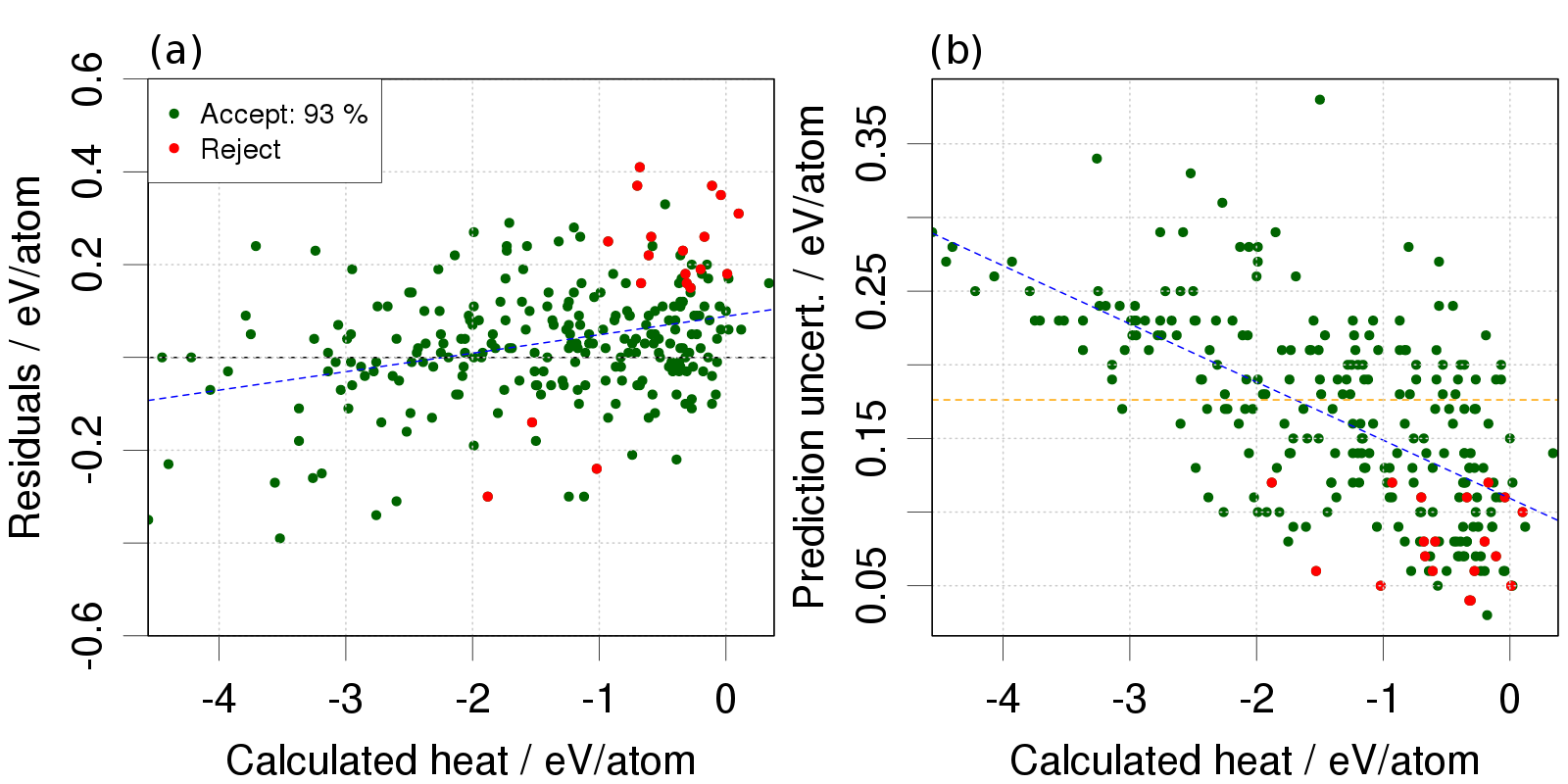} 
\par\end{centering}
\caption{\label{fig:Pandey}Residuals (a) and prediction uncertainties (b)
of the mBEEF density functional on a dataset of formation heats. The
red points are those for which a 95\% confidence interval around the
calculated value does not contain the reference value.}
\end{figure}

In the publications on mBEEF, one has only access to histograms of
the scaled errors, which do not enable us to appreciate the structure
of the prediction uncertainty for this method. Thanks to the formation
heat data provided in a recent article \cite{Pandey2015}, one can
now compare the prediction uncertainty with the residual errors of
the calibrated method and test their conformity. The dataset of residual
errors and prediction uncertainty used below has been extracted from
Table I of this article. The measurement uncertainty of the reference
data is not provided, but the typical experimental uncertainty on
formation heats has been reported to be well below 0.1~eV/atom \cite{Stevanovic2012}.

The residual errors are plotted in Fig.~\ref{fig:Pandey}(a) as a
function of calculated heats (the control variable): their distribution
presents a small positive linear trend, but the amplitude is weak,
and one would not gain much by an additional a posteriori correction.
One can therefore assume that the method is well calibrated and enables
to make predictions without significant bias (smaller than 0.1\,eV/atom)
within the calibration range. One can also see that the residual errors
are often much larger than the typical reference data uncertainty,
revealing model inadequacy.

Prediction uncertainties generated by the mBEEF method are plotted
in Fig.~\ref{fig:Pandey}(b): they display a marked negative linear
dependency with the control variable, with a correlation coefficient
of -0.63 and a ratio of about 3 between the extreme average values
(blue dashed line). This trend is not observable on the absolute values
of the residuals.

The mean prediction uncertainty $\overline{u}_{M|D}$ (0.18\,eV/atom)
is slightly higher than the $RMSE$ (0.14\,eV/atom). As a consistency
check, one calculates for each residual a $95$\,\% confidence interval
using the prediction uncertainty provided in the original article
and checks if this interval contains $0$. This is verified in $93$\,\%
of the cases, confirming the good average properties of the estimated
prediction errors. However, instead of being uniformly distributed
over the heat range, all the intervals failing the test appear only
for formation heats above $-2$\,eV/atom (red points in Fig.~\ref{fig:Pandey}),
\emph{i.e.}, $100$\,\% of the intervals below $-2$\,eV/atom include
the null value.

This leads us to conclude that the uncertainty of the lower heats
is overestimated (Fig.~\ref{fig:Pandey}(b)), while the uncertainty
of a fraction of the higher heats is underestimated. The $RMSE$ for
the heats below $-2$\,eV/atom is about 0.14\,eV/atom, while the
mean prediction uncertainty for this group is about 0.23\,eV/atom
($\sim$65\,\% overestimation).

Even if the effect is less striking than in the vibrational frequencies
case (Section \ref{sec:Vibrational-scaling-factors}), this example
shows also that indirect PUI produces prediction uncertainties that
are not distributed like the model errors they are supposed to represent.

\subsection{Lennard-Jones parameters\label{subsec:Lennard-Jones-parameters}}

The third example concerns the estimation of the $\sigma$, $\epsilon$
Lennard-Jones (LJ) potential parameters from the analysis of temperature-dependent
viscosity data. The data and Chapman-Enskog viscosity model are described
in Cailliez and Pernot \cite{Cailliez2011}. The reference set contains
41 points $\left\{ x_{i},\,y_{i},\,u_{y_{i}}\right\} $, where $x$
is the temperature, $y$ the viscosity and $u_{y}$ is the viscosity
measurement uncertainty. They result from 5 measurement series $\left\{ D^{(i)}\right\} _{i=1}^{5}$,
but one did not attempt here to model inter-series discrepancy. Therefore
the data covariance matrix is diagonal, with $\boldsymbol{V}_{D,ij}=u_{y_{i}}^{2}\delta(i-j)$
\cite{Cailliez2011,Pernot2017}.

The indirect and direct PUI methods presented above have been implemented
in \texttt{Stan} \cite{Gelman2015}, using the \texttt{rstan \cite{Rstan2016}}
interface package for \texttt{R} \cite{RTeam2015}. \texttt{Stan}
is a very flexible and efficient probabilistic programming language
to implement bayesian statistical models. A link to the codes to reproduce
the results of this example is provided in the Supporting Information
section. The indirect method (Sect.~\ref{par:Variance-inflation}),
implementing Eqns.~\ref{eq:likelihood-1} and \ref{eq:T-RB}, is
named VarInf\_Rb; the direct method based on marginal likelihoods
(Sect.~\ref{par:Model-level}) is named Margin; and the approximate
bayesian method (Sect.~\ref{par:Likelihood-level}) ABC. The covariance
matrix of the parameters is parameterized by $u_{\epsilon}$, $u_{\sigma}$
and $\rho$, the uncertainty on $\epsilon$, $\sigma$, and their
correlation coefficient, respectively. 

A \texttt{Stan} code provides a sample of the posterior pdf of the
model's parameters $p(\boldsymbol{\mu}_{\vartheta},\boldsymbol{V}_{\vartheta}|\boldsymbol{D})$,
from which statistics are calculated. The No-U-Turn sampler \cite{Hoffman2014}
was used, and convergence of the sampling was assessed by examining
the parameters samples and the \emph{split Rhat} statistics provided
by \texttt{rstan} \texttt{\cite{Rstan2016}}. Uniform prior pdfs have
been used for location parameters, and log-uniform for scaling parameters,
unless stated explicitly. All models were run with 4 parallel Markov
Chains of 5000 iterations each, 1000 of which are used as warm-up
for the No-U-Turn sampler. The convergence criteria and parameters
statistics are thus estimated on four samples of 4000 points. 

The mean values of the parameters and hyperparameters estimated for
all methods are reported in Table~\ref{tab:Ar}, along with their
Birge ratio ($R_{B})$, and mean prediction uncertainty $\overline{u}_{M|D}$.
The $RMSE$ for all methods is 0.1\,$\mu$Pa.s.

The Birge ratio for a model implementing the standard likelihood (Eq.~\ref{eq:likelihood})
is $R_{B}\simeq6.6$ (method WLS in Table~\ref{tab:Ar}), indicating
clearly that the Chapman-Enskog model is unable to fit the data within
their uncertainty range. Model inadequacy is also apparent through
the trend/oscillation in the residuals (Fig.~\ref{fig:Inadequacy}(b)).
As the VarInf\_Rb method has inflated data uncertainty (the scale
factor $T$ has been estimated from the Birge ratio of the WLS method
by Eq.~\ref{eq:T-RB}, giving $T\simeq129$), its Birge ratio (0.05)
is too small. The Margin method achieves a near-unit Birge ratio,
but the ABC method cannot reach this value, because of the constraints
introduced in the likelihood (Eq.~\ref{eq:ABC}).

\begin{table}[!t]
\caption{\label{tab:Ar}Parameters of the posterior pdf of the Lennard-Jones
parameters recovered by different PUI methods. The RMSE for all fits
is 0.10 $\mu$Pa.s.}

\noindent \centering{}%
\begin{tabular}{llr@{\extracolsep{0pt}.}lr@{\extracolsep{0pt}.}lr@{\extracolsep{0pt}.}lr@{\extracolsep{0pt}.}lr@{\extracolsep{0pt}.}lr@{\extracolsep{0pt}.}l}
\hline 
Method & $\mu_{\epsilon}$ & \multicolumn{2}{c}{$\mu_{\sigma}$} & \multicolumn{2}{c}{$u_{\epsilon}$} & \multicolumn{2}{c}{$u_{\sigma}$} & \multicolumn{2}{c}{$\rho$} & \multicolumn{2}{c}{$R_{B}$} & \multicolumn{2}{c}{$\overline{u}_{M|D}$}\tabularnewline
 & (K) & \multicolumn{2}{c}{($\text{\AA}$)} & \multicolumn{2}{c}{(K)} & \multicolumn{2}{c}{($\text{\AA}$)} & \multicolumn{2}{c}{} & \multicolumn{2}{c}{} & ($\mu$Pa&s)\tabularnewline
\hline 
WLS & 146.1(4) & 3&315(1) & \multicolumn{2}{c}{-} & \multicolumn{2}{c}{-} & -0&97 & 6&60 & 0&01\tabularnewline
VarInf\_Rb  & 146(5)  & 3&32(1)  & \multicolumn{2}{c}{- } & \multicolumn{2}{c}{- } & -0&97  & 0&05  & 0&13 \tabularnewline
Margin  & 146(1)  & 3&316(3)  & 0&6(8)  & 0&004(2)  & 0&0(6)  & 0&98  & 0&11 \tabularnewline
ABC  & 146.2(4)  & 3&315(1)  & 0&7(7)  & 0&003(2)  & 0&0(6)  & 1&20  & 0&09 \tabularnewline
Margin1  & 144(2)  & 3&321(4)  & 5&0(10)  & 0&015(3)  & -0&98(2)  & 0&88  & 0&13 \tabularnewline
Margin2  & 146(1)  & 3&315(3)  & 0&01(2)  & 0&0043(6)  & 0&0(6)  & 1&00  & 0&11 \tabularnewline
Margin3  & 145(1)  & 3&318(3)  & 1&6(2)  & 0&0001(2)  & 0&0(6)  & 0&98  & 0&10 \tabularnewline
\hline 
\end{tabular}
\end{table}

\begin{figure}[t]
\begin{centering}
\includegraphics[width=1\textwidth]{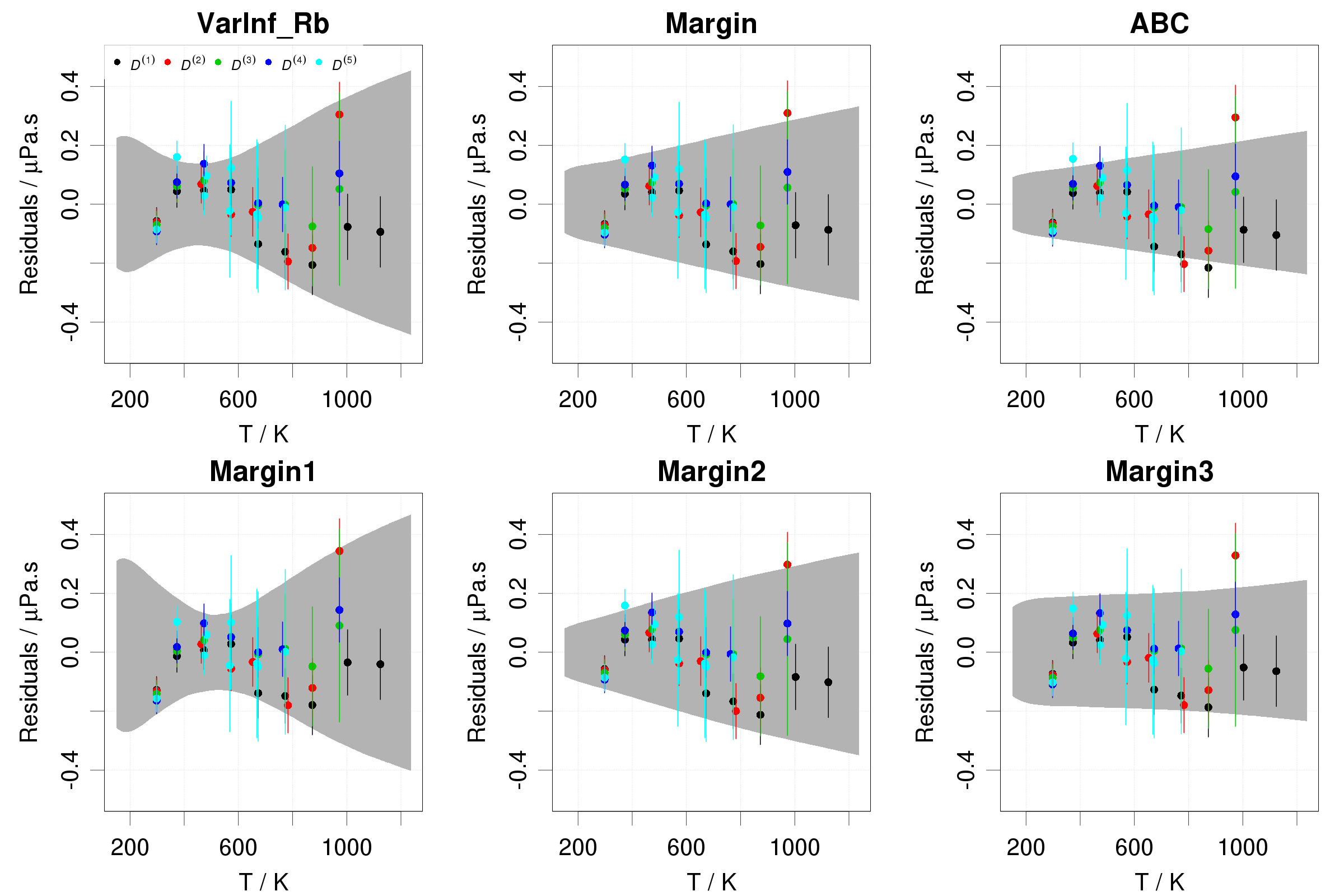} 
\par\end{centering}
\caption{\label{fig:Argon}Residuals and centered prediction bands of a Chapman-Enskog
model of Ar viscosity for the VarInf\_Rb, Margin and ABC methods (top
row), and for the three degenerate solutions of the Margin method
(bottom row). The dark-gray bands represent model prediction confidence
interval at the 2-$\sigma$ level, corrected from the mean prediction. }
\end{figure}

The residuals (points) and prediction band (gray area) for the indirect
method (VarInf\_Rb) are shown in Fig.~\ref{fig:Argon}. To be comparable
with the residuals, the prediction bands are corrected from the mean
prediction value, $\mu_{M|D}(T)$. One sees that the prediction band
adopts a diabolo structure, with a pronounced waist around 500\,K.
By contrast, optimization of the covariance matrix of the parameters
by the direct methods, Margin and ABC, leads to prediction bands with
more regular shapes (Fig.~\ref{fig:Argon}). All methods achieve
a mean prediction uncertainty in fair agreement with their $RMSE$
of 0.1\,$\mu$Pa.s, although VarInf\_Rb returns a value slightly
in excess, with $\overline{u}_{M|D}=0.13$\,$\mu$Pa.s. It is difficult
at this stage to pick a best prediction uncertainty model: the Margin
method might be favored due to its better Birge ratio.

\begin{figure}[!t]
\noindent \begin{centering}
\includegraphics[width=0.9\textwidth]{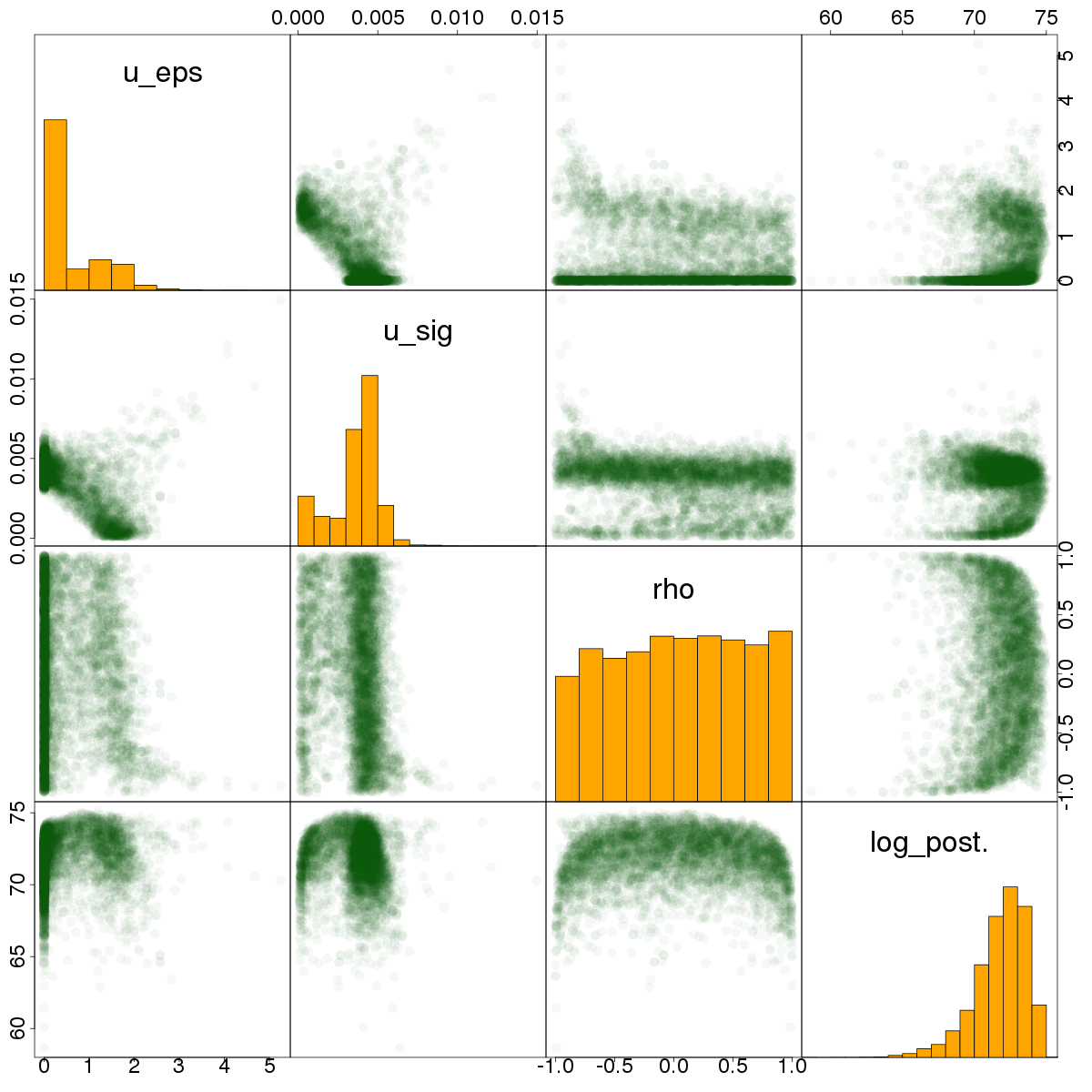} 
\par\end{centering}
\caption{\label{fig:Scatterplots}Scatterplots matrix of the posterior sample
for the Margin method's covariance matrix parameters, showing its
multimodality; the diagonal provides histograms of the parameters
posterior distribution, while the out-of-diagonal plots represent
a projection of the sample in the 2D space of the corresponding parameters
pair (the upper and lower matrices are redundant); ``log\_post.''
is the logarithm of the posterior pdf.}
\end{figure}

Inspection of a sample of the posterior pdf for the Margin (Fig.~\ref{fig:Scatterplots})
and ABC methods (not shown) reveals the presence of three modes (high-density
areas), each one corresponding to a minimum value of a parameter of
the $\boldsymbol{V}_{\vartheta}$ covariance matrix. In the present
case, the mode corresponding to $\rho\simeq-1$ is less outstanding
than the modes at $u_{\epsilon}\simeq0$ and $u_{\sigma}\simeq0$. 

By constraining the support of the parameters through their prior
pdf, the three modes have been sampled independently for the Margin
method. They are reported as Margin1 to Margin3 in Table~\ref{tab:Ar},
and in Fig.~\ref{fig:Argon}. The three samples produce very slightly
different estimates of the LJ parameters, achieve good Birge ratios
near unity, but present marked differences on the variance matrix
parameters: 
\begin{itemize}
\item Margin1 corresponds to an extreme negative correlation of $\epsilon$
and $\sigma$ , and to large values of both $u_{\epsilon}$ and $u_{\sigma}$.
This solution gives a prediction band very similar to the one of VarInf\_Rb
(Fig.~\ref{fig:Argon}), and leads to the same excess in mean prediction
uncertainty (Table~\ref{tab:Ar}). Its Birge ratio value (0.88) is
the smallest of the 3 modes. 
\item Margin2 corresponds to a minimal value of $u_{\epsilon}$ and an undetermined
value of $\rho$. Considering the prediction bands the Margin2 mode
appears to have a major contribution to the Margin sample.
\item Margin3 is the symmetrical of Margin2, with a minimal value of $u_{\sigma}$,
and corresponds to an almost uniform prediction band. 
\end{itemize}
The direct PUI methods Margin and ABC are therefore subject to a degeneracy
in the optimal hyperparameters describing the stochastic LJ parameters,
the implications of which are presented in the next section.

\section{Discussion and Conclusion\label{sec:Discussion-and-Conclusion}}

We have presented PUI methods in a unified formalism and established
links between these methods, and with other methods in uncertainty
quantification. We have shown through a series of examples in different
contexts that existing methods attempting to capture model inadequacy
errors in the covariance matrix of the model parameters present a
series of problematic properties.

The prediction uncertainty bands are constrained by the functional
form of the sensitivity coefficients of the model (Eq.~\ref{eq:Sensitivity}),
notably when using a simple inflation of the data variance (indirect
method, Sect.~\ref{par:Variance-inflation}). This has been apparent
in the three examples above, and it might lead to areas of the control
parameter with systematic under- or over-estimation of the prediction
uncertainty. Unfortunately, this information is hard to gather directly
from the literature, as the authors report typically the mean prediction
uncertainty, or histograms of prediction uncertainties, which mask
trends or systematic effects along the control variable. We can only
recommend that authors working with these methods provide more informative/detailed
representations of prediction uncertainties, and discuss the impact
of under- or over-estimated prediction uncertainties on their intended
use.

The influence of the model sensitivity coefficients can be modulated
by the covariance matrix of the parameters, when the latter is optimized
(direct methods, Sect.~\ref{subsec:Direct-approach}). We have seen
in Sect.~\ref{subsec:Lennard-Jones-parameters}, that direct PUI
methods based on a stochastic representation of the model's parameters
might present degenerate modes leading to very different shapes of
the prediction uncertainty bands, and that one has no \emph{a priori}
criterion to choose among them.\footnote{Note that when the reference data are abundant, which was not the
case in the chosen example, keeping aside a validation set might help
in this regard.} However, it is interesting that one of the modes of the Margin method
is similar to the solution obtained by the scaling of data variance,
and that this mode achieves sub-optimal statistics, both for its Birge
ratio and for its mean prediction uncertainty. This would suggest
that the indirect PUI method does not provide the best solution to
the prediction uncertainty estimation problem. As discussed by Pernot
and Cailliez \cite{Pernot2017}, the posterior pdf multimodality/degeneracy
might imply an undesirable high sensitivity of the prediction band
shape to changes in the calibration dataset. Besides, the multimodality
problem of the posterior pdf can be expected to increase with the
number of parameters.

Considering the direct PUI method, the Margin method has no tuning
option that would enable a performance improvement. At the opposite,
the empirical likelihood on which the ABC method is based enables
to envision additional constraints which might help to relieve the
multimodality problem. This is in our opinion the most promising route
to a design a satisfying PUI method. But, we have also seen that these
constraints tend to produce sub-optimal residuals, leading to a compromise
between fit quality and prediction uncertainty quality.

Linear uncertainty propagation (LUP) has been used to estimate the
mean value and covariance matrix of the model predictions in the Margin
and ABC methods (Eqns. \ref{eq:VM}, \ref{eq:residMean} and \ref{eq:ABC}).
In the present application (Ar viscosity), the uncertainty on the
model's parameters is small (less than 1\%) and the viscosity model
is monotonous and continuous in the LJ parameters variation range.
There is no reason to be worried about uncontrolled non-linearity
effects. However, this is not necessarily the case for other models,
and the use of LUP has to be handled with care. For instance, Pernot
and Cailliez \cite{Pernot2017} validated the use of LUP in a similar
application by estimating the relative errors between an LJ parameter-wise
linear approximation of the viscosity model over the whole T range
and a sample of model values for LJ parameters drawn from their posterior
pdf.

\medskip{}
 The main conclusion of this study is that methods to estimate prediction
uncertainty of inadequate models based on parameters uncertainty inflation
have to be used with great care and subjected to careful inspection,
both of parameter space, and of prediction uncertainty trends. There
is no sense in using prediction uncertainties if they are not reliable.

\section*{Supplementary Material}

See \href{https://zenodo.org/badge/latestdoi/88050574}{supplementary material}
for the data and codes used in Sections \ref{sec:Vibrational-scaling-factors}
and \ref{subsec:Lennard-Jones-parameters}.

\bibliographystyle{aipnum4-1}

\end{document}